
\input phyzzx

\overfullrule=0pt

\def\d{{\rm d}}
\def\half{{1 \over 2}}
\def\ra{\rightarrow}
\def\ve{\varepsilon}
\def\Lag{{\cal L}}
\def\M {{\cal M}}
\def\del{\partial}
\def\delslash{\rlap{/}\partial}
\def\bpsi{\overline{\psi}}
\def\Sp{{\rm Sp}}
\def\Re{{\rm Re~}}
\def\SU{{\rm SU}}
\def\Sigdag{ \Sigma^\dagger }

\def\ss{\sigma_s}
\def\st{\sigma_t}

\def\scale{ 2 \sqrt{2\pi} f }
\nopubblock
\line{\hfil JHU-TIPAC-920012}
\line{\hfil June, 1992}
\titlepage
\title{ Inelastic Channels in the Electroweak Symmetry-Breaking Sector }
\author{ S.~G.~Naculich \ \ and\ \ C.--P. Yuan}
\address{
Department of Physics and Astronomy \break
The Johns Hopkins University  \break
Baltimore, MD  21218 }

\abstract{It has been argued that
if light Higgs bosons do not exist
then the self--interactions of $W$'s
become strong in the TeV region and can be
observed in longitudinal $WW$ scattering.
We present a model with many inelastic channels
in the $WW$ scattering process,
corresponding to the creation of heavy fermion pairs.
The presence of these heavy fermions
affects the elastic scattering of $W$'s
by propagating in loops,
greatly reducing the amplitudes
in some charge channels.
Consequently, the symmetry--breaking sector cannot
be fully explored by using,
for example, the $W^+W^+$ mode alone;
all $WW \ra WW$ scattering modes must be measured.}

\endpage

\REF\nolose{
M. S. Chanowitz and M. K. Gaillard, Nucl. Phys. {\bf B261}, 379 (1985);
\nextline
M. S. Chanowitz, Ann. Rev. Nucl. Part. Sci. {\bf 38}, 323 (1988).}
\REF\wpwp{
M.~S.~Chanowitz and M.~Golden, Phys.\ Rev.\ Lett.\ {\bf 61}, 1053
(1988); {\bf 63}, 466(E) (1989);
M.~S.~Berger and M.~S.~Chanowitz, LBL-30476 (1991).}
If light Higgs bosons do not exist, it is believed that
elastic longitudinal $WW$ scattering will be enhanced,
indicating the presence of new strong interactions
at or above 1 TeV.  (We use $W$ to denote either the $W^\pm$ or $Z^0$ boson.)
It has been claimed that the energy and luminosity of the
SSC are large enough that, whatever form the new interactions take,
they would be observable in $WW$ two--body interactions via
leptonic decays of $W$'s.\refmark{\nolose}
This is called the ``no--lose theorem.''
Study of the $W^+ W^+$ mode has been particularly promoted\refmark\wpwp\
in the context of observing these strong interactions
because the standard model background for this mode is small.

\REF\CG{
R. S. Chivukula and M. Golden, Phys. Lett. {\bf B267}, 233 (1991). }
\REF\hidden{
S.~G.~Naculich and C.--P.~Yuan, preprint JHU--TIPAC--920017.}
Recently, Chivukula and Golden have emphasized
the possible existence of inelastic channels in
the $WW$ scattering process.\refmark{\CG}
They studied an $O(4) \times O(n)$ model in which the $WW$ interactions are
almost entirely inelastic  (to $n$ species of pseudo--Goldstone bosons).
Consequently, the elastic $WW$ amplitude is reduced,
the more so for a larger number $n$ of inelastic channels.
(The total event rate for elastic scattering,
however, does not decrease as $n$ increases.\refmark{\hidden})
In that model, inelastic scattering
leads to a very broad low--energy resonance
in the elastic $WW$ scattering amplitudes.
In this letter, we present a model with many inelastic channels
in the electroweak sector and which has no resonances.

In general, the presence of inelastic channels
corresponds to additional particles in the theory.
Even if the production of these particles
is not directly observed,
they affect the {\it elastic} scattering of $W$'s by
propagating in loops.
These loops necessarily contribute to
the imaginary part
of the elastic scattering amplitude,
which is related by the optical theorem
to the total cross section.
The loops also contribute to
the real part
of the elastic amplitude,
interfering with the Born contribution.
This interference may dramatically
reduce the signal in some charge channels,
{\it e.g.} the $W^+W^+$ channel.
Other channels may be enhanced, however,
both by real and imaginary loop corrections.
The model we present below
has precisely this behavior.
The moral is that
to be certain of detecting the symmetry--breaking sector
it will be necessary to measure scattering
in all the final state $WW$ modes.

The no--lose theorem,
with its prediction of strong $WW$ scattering
in the absence of a light Higgs resonance,
is based on the low--energy theorems for
a theory with spontaneously--broken symmetry.
The pattern of symmetry breaking for
the electroweak sector is
$ \SU(2)_L \times  \SU(2)_R \longrightarrow \SU(2)_V$
if we assume that
it respects a custodial SU(2) symmetry.
At energies $s \gg M_W^2$,
the longitudinal vector bosons $W$
correspond,
via the equivalence theorem,
to the three Goldstone bosons $\phi^a$
resulting from this broken symmetry.
The interactions of these Goldstone bosons
are described by low--energy theorems,
which emerge automatically
when we describe this broken symmetry
using chiral Lagrangians.
In this approach,
the Goldstone bosons
are parametrized by the matrix
$$
\Sigma
=\exp{\left( i \tau^a \phi^a \over f \right) }, \quad\quad a=1,2,3,
\eqn\sigmafield
$$
where $\tau^a$ are the Pauli matrices.
The lowest energy term of the chiral Lagrangian is
$$
\Lag = {f^2 \over 4}
{\rm Tr} \left(\del_\mu \Sigma \del^\mu \Sigdag \right).
\eqn\LETLag
$$
It follows that the low--energy amplitude for
$ Z^0 Z^0 \to W^+ W^-$, for example,
is
$$
\M (Z^0Z^0 \to W^-W^+)  =  {s \over  f^2},
\eqn\LET
$$
where
$f=250$ GeV.
This amplitude grows with
increasing center--of--mass energy,
becoming strong at around 1 TeV.

\REF\BDV{
J.~Bagger, S.~Dawson and G.~Valencia, preprint BNL-45782;
preprint JHU-TIPAC-920009.}
At higher energies,
the amplitude \LET~is modified by
corrections of order
$ s^2 / 16\pi^2 f^4 $
coming from higher--dimension operators
induced by physics at a higher scale,
and from Goldstone boson loop corrections.\refmark{\BDV}
Indeed, we know that the amplitude \LET~must
eventually break down,
because it violates partial--wave unitarity
($ |{\rm Re}~a_0^0|  > \half$)
at about $\scale \sim 1.2$ TeV.
Nevertheless,
below the scale of unitarity breakdown,
the scattering of $W$'s
will be roughly governed by eq.~\LET,
as long as there are
no ``light'' particles
($ M \lsim 1$ TeV)
in the symmetry--breaking sector
other than the Goldstone bosons.
On the other hand,
if there do exist such light particles,
the amplitude \LET~will
only be valid for $ s \ll M^2 $.

If such light particles can be exchanged by the $W$'s,
as in the case of a light Higgs boson for example,
there will be narrow resonances in the $WW$ scattering amplitudes
at $s \sim M^2$.
We consider another possibility,
particles with mass well below 1 TeV
which can only be produced in pairs;
in particular we have in mind heavy fermions.
These fermions correspond to inelastic channels in the
$WW$ scattering process.
Unlike exchange particles,
they will not necessarily produce resonances
in the elastic $WW$ scattering amplitudes.
They alter these amplitudes
through loop effects, however,
and lead to behavior
markedly differently from that
predicted by low--energy theorems
at scales above
the threshold for fermion pair production
but below 1 TeV.
(New physics must still enter at around 1 TeV,
because the fermions,
unlike the Higgs boson,
do not unitarize the amplitudes.)

In this letter,
we present a chiral Lagrangian model
coupled to heavy fermion doublets.
This model has no resonances in the
elastic $WW$ scattering amplitudes,
therefore we must study these amplitudes
in the TeV region where they become strong.
We find that the presence of the fermions
dramatically reduces the scattering rates
for the $W^+ W^+$ and $W^+ Z^0$ modes
relative to the predictions of
the low--energy theorems.
On the other hand,
the rates for scattering into the modes
$W^+ W^-$ and $Z^0 Z^0$ are enhanced.
This enhancement partially results
from the large imaginary part
of the loop amplitude in the forward direction,
which via the optical theorem
is due to the large total cross section.
Here we will only present our results; details of the
calculation will be given
elsewhere.\Ref\ny{S. G. Naculich and C.--P. Yuan, in preparation.}

The Lagrangian for the model is
$$
\eqalign{
\Lag
&
=
{N v^2 \over 4}
{\rm Tr} \left(\del_\mu \Sigma \del^\mu \Sigdag \right)
+ \sum^{N}_{j=1} \bigg[ \bpsi^j  ~i\delslash ~ \psi^j
-gv  \left( \bpsi_L^j \Sigma  \psi_R^j
          + \bpsi_R^j \Sigdag \psi_L^j \right) \bigg],
\cr
 \Sigma
&
=\exp{\left( i \tau^a \phi^a \over \sqrt{N} v \right) }~,
 \quad\quad
\psi_L^j  = {1\over 2} \left( 1 - \gamma_5 \right) \psi^j ~,
 \quad\quad
\psi_R^j  = {1\over 2} \left( 1 + \gamma_5 \right) \psi^j ~.
\cr}
\eqn\FermLag
$$
The fields $\psi^j$
represent $N$ degenerate heavy fermion doublets
with mass $m=gv$.
The effects of the fermions
on the $WW$ scattering amplitudes
will be significant when the Yukawa coupling $g$ is large.
To capture this,
we will not calculate the amplitudes
perturbatively in $g$,
but rather in a $1/N$ expansion,
holding the parameters
$g$ and $v$ fixed as $N \to \infty$.
The results will be valid for arbitrary Yukawa coupling $g$,
\ie, for all values of the fermion mass $m$.

Were we to calculate perturbatively in the Yukawa coupling,
the real part of the loop correction would contribute
through interference with the tree--level amplitude,
but the imaginary part would be higher order.
In the large--$N$ approach,
the imaginary part
of the loop correction
contributes in leading order.
Because it is related to the total cross section
via the optical theorem,
this contribution is important
when there are many inelastic channels.

To leading order in $1/N$,
the only corrections
come from fermion loops.
These contribute a divergence to the
Goldstone boson self-energy.
Accordingly, we add a counterterm
$$
\delta \Lag
= \delta Z \left( N v^2 \over 4 \right)
{\rm Tr} \left(\del_\mu \Sigma \del^\mu \Sigdag \right),
\eqn\Counter
$$
with $\delta Z$ chosen so that the residue of the
Goldstone boson propagator at $p^2 = 0$ is unity.
This prescription will ensure that the low--energy theorems
for the scattering amplitudes are satisfied.

The Lagrangian \FermLag~is {\it not} the most general one
with global $ \SU(2)_L \times  \SU(2)_R$ chiral symmetry;
we have omitted a possible derivative coupling of the form
$  \kappa_L \bpsi_L  (\Sigma i \delslash  \Sigdag) \psi_L
+  \kappa_R \bpsi_R  (\Sigdag i \delslash  \Sigma) \psi_R $.
(If parity is conserved, then $\kappa_L = \kappa_R$.)
We also have not included any four-derivative terms
involving  $\Sigma$.
To leading order in $1/N$,
no such terms are needed to
absorb divergences;
the counterterm \Counter~suffices to
cancel the divergences
of the fermion loops
in the $WW \ra WW$ amplitudes.

The $WW$ scattering amplitudes
are given by
$$
\eqalign{
\M (Z^0Z^0 \to W^-W^+) & = A(s,t,u), \cr
\M (W^-W^+ \to Z^0Z^0) & = A(s,t,u), \cr
\M (W^-W^+ \to W^-W^+) & = A(s,t,u)+A(t,s,u), \cr
\M (Z^0Z^0 \to Z^0Z^0) & = A(s,t,u)+A(t,s,u)+A(u,t,s), \cr
\M (W^\pm Z^0 \to W^\pm Z^0) & = A(t,s,u), \cr
\M (W^\pm W^\pm \to W^\pm W^\pm) & =A(t,s,u)+A(u,t,s). \cr}
\eqn\threefive
$$
Including only diagrams
which contribute to leading order in $1/N$,
we find
$$
A(s,t,u)
 = {1\over N} \bigg\{
{s\over v^2} - {m^2 \over 4 \pi^2 v^4} s F_2 (s)
- {m^4 \over 4 \pi^2 v^4}
\Big[ F_4 (s,t)  + F_4 (s,u)  -  F_4 (t,u)  \Big] \bigg\},
\eqn\fourthree
$$
where
$$
\eqalign{
F_2 (s)
& = \int_0^1 \d x
\ln \left[ 1 - {s\over m^2} x(1-x) - i \ve \right],
\cr
F_4 (s,t)
&
= \int_0^1  \d x
\left[ x^2 - x + { m^2(s+t)\over st } \right]^{-1}
\cr
&
{}~~~~\times \biggl\{ \ln \Bigl[ 1 - {s\over m^2} x(1-x) - i \ve \Bigr]
+       \ln \Bigl[ 1 - {t\over m^2} x(1-x) - i \ve \Bigr]
\biggr\}.
\cr}
\eqn\fourfour
$$
The integral $F_2(s)$ is given by
$$
\eqalign{
F_2(s<0)
&  =  -~2+~\sqrt{ 1 - {4m^2 \over s} }
  ~\ln\left(
  { \sqrt{4m^2-s} + \sqrt{-s}  \over
    \sqrt{4m^2-s} - \sqrt{-s}  } \right) ,
\cr
F_2( 0<s<4m^2)
& =-~2+2~\sqrt{ -1 + {4m^2 \over s}  } ~{\rm arctan}
  \sqrt{ {s \over 4m^2-s} },
\cr
F_2(s>4m^2)
& =-~2+\sqrt{ 1 - {4m^2 \over s}}
\left[ \ln\left(
 { \sqrt{s} + \sqrt{s-4m^2} \over   \sqrt{s} -  \sqrt{s-4m^2} }
\right) - i  \pi\right],
\cr}
\eqn\threeseven
$$
and the integral $F_4 (s,t)$ can be written\Ref\spence{
G. 't Hooft and M. Veltman, Nucl.~Phys.~{\bf B153}, 365 (1979).}
in terms of Spence functions as
$$
\eqalign{
F_4&(s,t)
= 2 \left[ 1 - { 4m^2(s+t)\over st } \right]^{-\half}
\bigg[ \Sp \left( {x_+  \over x_+ - y_+(s) } + i\ss\ve \right)
\cr
&
     + \Sp \left( {x_+  \over x_+ - y_-(s) } - i\ss\ve \right)
     - \Sp \left( {-x_- \over x_+ - y_+(s) } + i\ss\ve \right)
     - \Sp \left( {-x_- \over x_+ - y_-(s) } - i\ss\ve \right)
\cr
&
     + \Sp \left( {x_+  \over x_+ - y_+(t) } + i\st\ve \right)
     + \Sp \left( {x_+  \over x_+ - y_-(t) } - i\st\ve \right)
     - \Sp \left( {-x_- \over x_+ - y_+(t) } + i\st\ve \right)
\cr
&
     ~~~~~~~~~~~~~~~~~~~~~
     - \Sp \left( {-x_- \over x_+ - y_-(t) } - i\st\ve \right)
     + 2\pi i \Theta( st ) \ln \left( x_+ \over - x_- \right)
     \bigg],
\cr}
\eqn\fourfive
$$
where $\ss$ denotes the sign of $s$, and
$$
x_\pm = \half \pm \sqrt{ {1\over 4} - {m^2 (s+t)\over st} } ~,
{}~~~~~~~
y_\pm (s) = \half \pm \sqrt{ {1\over 4} - {m^2 \over s} } ~.
\eqn\foursix
$$
Now we fix $N$ to a finite value,
the number of fermion doublets in the model.
Then we set the scale $v$ equal to $f/\sqrt{N}$,
where $f = 250$ GeV characterizes the scale of symmetry breaking,
to obtain
$$
A(s,t,u)
=   {s\over f^2}
  - {N m^2 \over 4 \pi^2 f^4} s F_2 (s)
  - {N m^4 \over 4 \pi^2 f^4}
\Big[ F_4 (s,t)  + F_4 (s,u)  -  F_4 (t,u)  \Big].
\eqn\ampfunc
$$
In the limit $s \ll m^2$,
$ A(s,t,u) $ approaches $ s/f^2 $,
in accord with low--energy theorems.

Our model depends on only two parameters:
the number of fermion doublets $N$ and the fermion mass $m$.
Experimental bounds require $N$ to be $ \le 15$.
This constraint derives from the
contribution of additional heavy degenerate fermions
to the shift in $M_W$.\Ref\wmass{
M. Veltman, Phys. Lett. {\bf 91B}, 95 (1980);\nextline
W. J. Marciano and A. Sirlin, Phys. Rev. {\bf D22}, 2695 (1980);\nextline
S. Bertollini and A. Sirlin, Nucl. Phys. {\bf B248}, 589 (1984).}

We consider a model with $m = 250$ GeV and $N=15$,
corresponding to 5 additional generations
of degenerate quark doublets,
due to the color degeneracy.
For these parameters,
the partial waves obey unitarity
($ | a_I^J | \le 1$)
for the energy region which we will consider.\Ref\unit{
The partial waves in both our model (with these parameters)
and the low--energy theorem model
satisfy $ |\Re a_I^J| \le \half$
up to about 1.2 TeV.}
In table~1,
we show the event rates and angular distributions
of the $W$'s produced in various modes at the SSC
(with integrated luminosity $10^4$  pb$^{-1}$)
for this model,
comparing them with results\Ref\CG{
We use the scattering amplitudes given by
M. S. Chanowitz and M. K. Gaillard,
cited in ref.~\nolose.}
from the ``low--energy theorem model,''
with Lagrangian \LETLag.
We do not include $W$ pairs
produced by either quark or gluon fusion,
restricting our consideration to $WW$ scattering.
The invariant mass of the $W$ pair is required to be
within 850 GeV and 1350 GeV.
The branching ratio of the $W$ boson decay has not been included.
The transverse momentum of each decay product of the $W$'s
is required to be at least 20 GeV.
No rapidity cut has been imposed on the final--state particles.
The event rate is calculated using the effective--$W$ approximation.
The parton distribution function used is the leading order set,
Fit SL, of Morfin and Tung.\Ref\MT{
J. G. Morfin and Wu--Ki Tung, Z. Phys. {\bf C52}, 13 (1991).}
The scale used in evaluating the parton distribution function
in conjunction with the effective--$W$ method is $M_W$.

{}From the results of table~1,
we see that
in the chiral Lagrangian model
with heavy fermion doublets,
the rates for the $W^+Z^0$ and $W^+W^+$ modes
are greatly reduced
relative to the low--energy theorem model.
In fact,
the $W^+W^+$ event rate at $M_{WW}=1.5$ TeV
is down by about a factor of 10
from the prediction of the low--energy theorem model,
and the $W^+Z^0$ mode is reduced by a similar factor.
This is because
the tree level amplitude is largely cancelled
by the real part of the loop amplitude,
and the imaginary part of the amplitude
is zero for $W^+W^+$
and small for $W^+Z^0$.
On the other hand,
the heavy fermion model
has a larger rate
than the low--energy theorem model
in both the $W^-W^+$ and $Z^0Z^0$ modes
because of enhancement from the loop contribution to the scattering
amplitudes.
This increase, however,
is less than a factor of 2
even at $M_{WW}=1.5$ TeV.
The event rate for the $W^-W^+$ and $Z^0Z^0$ modes
in the TeV region
is almost entirely due to
the imaginary forward part of the amplitude
(related by the optical theorem to the total cross section),
the effect becoming stronger for higher $M_{WW}$.

In this letter,
we have examined the effects
of inelastic channels
in the $WW$ scattering process
in one specific model.
In this model,
containing heavy fermion doublets,
the rate of elastic $WW$ scattering
at energies above the threshold for fermion pair production
differs significantly
from low--energy theorem predictions.
In particular,
we found a large suppression
of some elastic $WW$ charge channels
and an enhancement of others.
A lesson to be drawn from this model
is that all charge modes of the $WW \ra WW$ process
need to be observed
to be sure of detecting the symmetry--breaking sector.

\ack

It is a pleasure to thank G.~L.~Kane
for asking the questions which stimulated this work,
and for many fruitful discussions.
We are also grateful to
J.~Bagger,
J.~Bjorken,
Gordon Feldman,
B.~Grinstein,
C.~Im,
G.~Ladinsky,
S.~Meshkov,
S.~Mrenna,
F.~Paige,
and
E.~Poppitz
for discussions.
This work has been supported by the National Science Foundation
under grant no.~PHY-90-96198.

\endpage
\refout

\endpage
\singlespace
\tabskip=1em plus2em minus.5em
\newdimen\digitwidth\setbox0=\hbox{$0$}\digitwidth=\wd0\catcode`?=\active
\def?{\kern\digitwidth}
\newdimen\digitwidtha\setbox0=\hbox{$\le$}\digitwidtha=\wd0\catcode`@=\active
\def@{\kern\digitwidtha}
\twelverm
\null
\vskip0.5cm
\vbox{
$$\vbox{
\def\tablerule{\noalign{\smallskip\hrule\vskip .04in \hrule \smallskip}}
\twelvepoint
\halign to \hsize {\hfil#\hfil&\hfil#\hfil&\hfil#\hfil&
\hfil#\hfil\cr
\multispan3\hfil TABLE~1 \hfil\cr
\tablerule
$(W^+W^+)$    & LET & HF \cr
\noalign{\smallskip\hrule\medskip}
$ 0 < |\eta| < 1$  &  94 & 22 \cr
$ 0 < |\eta| < 2$  & 246 & 57 \cr
$ 0 < |\eta| < 4$  & 343 & 80 \cr
others             &   3 &  1 \cr
\noalign{\smallskip \hrule}}}$$}
\vbox{
$$\vbox{
\def\tablerule{\noalign{\smallskip\hrule\vskip .04in \hrule \smallskip}}
\twelvepoint
\halign to \hsize {\hfil#\hfil&\hfil#\hfil&\hfil#\hfil&
\hfil#\hfil\cr
\tablerule
$(W^-W^+)$         & LET &  HF \cr
\noalign{\smallskip\hrule\medskip}
$ 0 < |\eta| < 1$  & 199 & 333   \cr
$ 0 < |\eta| < 2$  & 530 & 890   \cr
$ 0 < |\eta| < 4$  & 741 & 1257  \cr
others             &   6 &  11   \cr
\noalign{\smallskip \hrule}}}$$}
\vbox{
$$\vbox{
\def\tablerule{\noalign{\smallskip\hrule\vskip .04in \hrule \smallskip}}
\twelvepoint
\halign to \hsize {\hfil#\hfil&\hfil#\hfil&\hfil#\hfil&
\hfil#\hfil\cr
\tablerule
$(Z^0Z^0)$         & LET & HF \cr
\noalign{\smallskip\hrule\medskip}
$ 0 < |\eta| < 1$  & 119 &  167 \cr
$ 0 < |\eta| < 2$  & 308 &  444 \cr
$ 0 < |\eta| < 4$  & 421 &  622 \cr
others             &   3 &    6 \cr
\noalign{\smallskip \hrule}}}$$}
\vbox{
$$\vbox{
\def\tablerule{\noalign{\smallskip\hrule\vskip .04in \hrule \smallskip}}
\twelvepoint
\halign to \hsize {\hfil#\hfil&\hfil#\hfil&\hfil#\hfil&
\hfil#\hfil\cr
\tablerule
$(W^+Z^0)$    	   & LET & HF \cr
\noalign{\smallskip\hrule\medskip}
$ 0 < |\eta| < 1$  &  88 &  22  \cr
$ 0 < |\eta| < 2$  & 253 &  69  \cr
$ 0 < |\eta| < 4$  & 373 & 110  \cr
others        	   &   4 &   2  \cr
\noalign{\smallskip \hrule}}}$$}

\vbox{
\Tenpoint
\noindent
TABLE~1.
The event rates for
$W^+W^+$, $W^-W^+$, $Z^0Z^0$, and $W^+Z^0$
production in one SSC year for the low--energy theorem model
and for the chiral Lagrangian model with heavy fermion doublets.
The invariant mass of the $W$ pair is required to satisfy
$850 \, {\rm GeV} < M_{WW} < 1350$ GeV.
The branching ratio of the $W$ boson decay is not included.
$ 0 < |\eta| < 1$ means that both
$W$ bosons have pseudo--rapidity between 0 and 1, {\it etc.} }

\end